\documentclass[conference]{IEEEtran}
\IEEEoverridecommandlockouts
\usepackage{balance}
\usepackage{cite}
\usepackage{amsmath,amssymb,amsfonts}
\usepackage{algorithmic}
\usepackage{graphicx}
\usepackage{textcomp}
\usepackage{xcolor}
\usepackage[nolist]{acronym}
\def\BibTeX{{\rm B\kern-.05em{\sc i\kern-.025em b}\kern-.08em
    T\kern-.1667em\lower.7ex\hbox{E}\kern-.125emX}}
\begin{document}

\title{\fontsize{19}{21}\selectfont
FADE: A Flow-Based Approach for Multichannel Tail Estimation
}

\author{
    \IEEEauthorblockN{
    Parmida Valiahdi\IEEEauthorrefmark{1}, 
    Niloofar Mehrnia\IEEEauthorrefmark{2},
    Mustafa Furkan Beker\IEEEauthorrefmark{3}\\[2pt]
    Ubeydullah Erdemir\IEEEauthorrefmark{3}\IEEEauthorrefmark{4},
    Ibrahim Hokelek\IEEEauthorrefmark{3}\IEEEauthorrefmark{4},
    Sinem Coleri\IEEEauthorrefmark{1}}
    
    \IEEEauthorblockA{\IEEEauthorrefmark{1}Department of Electrical and Electronics Engineering, Koc University, Istanbul, Turkey}
    \IEEEauthorblockA{\IEEEauthorrefmark{2}Department of Electrical Engineering and Computer Science, KTH Royal Institute of Technology, Stockholm, Sweden}
    \IEEEauthorblockA{\IEEEauthorrefmark{3}Communications and Signal Processing Research (HISAR) Lab., TUBITAK-BILGEM, Kocaeli, Turkey}
    \IEEEauthorblockA{\IEEEauthorrefmark{4}Department of Electronics and Communication Engineering, Istanbul Technical University, Istanbul, Turkey}

\thanks{Sinem Coleri acknowledges the support of TUBITAK (Scientific and Technological Research Council of Türkiye) Science Award grant and TUBITAK  2247-A National Leaders
Research Grant \#121C314. Parmida Valiahdi acknowledges the support of TUBITAK BiDEB 2211-A National PhD Scholarship Program. Niloofar Mehrnia acknowledges the support of  Vinnova (Sweden’s Innovation Agency) under Grant \#2025-01333.}
}

\maketitle
\begin{acronym}
    \acro{URLLC}{Ultra-Reliable Low Latency Communication}
  \acro{EVT}{Extreme Value Theory}
  \acro{GPD}{Generalized Pareto Distribution}
  \acro{RMSE}{Root Mean Square Error}
  \acro{MAE}{Mean Absolute Error}
  \acro{MLE}{Maximum Likelihood Estimation}
  \acro{GAN}{Generative Adversarial Network}
  \acro{EVT-HGAN}{EVT-guided Hybrid Generative Adversarial Network}
  \acro{VAE}{Variational Autoencoder}
  \acro{MLP}{Multilayer Perceptron}
  \acro{POT}{Peak-Over-Threshold}
  \acro{GEV}{Generalized Extreme Value}
  \acro{QQ}{Quantile-Quantile}
  \acro{KS}{Kolmogorov-Smirnov}
  \acro{PPCC}{Probability Plot Correlation Coefficient}
  \acro{CDF}{Cumulative Distribution Function}
  \acro{MSE}{Mean Squared Error}
  \acro{GRU}{Gated Recurrent Unit}
  \acro{LSTM}{Long Short-Term Memory}
  \acro{RNN}{Recurrent Neural Network}
  \acro{DRNN}{Deep Recurrent Neural Network}
  \acro{Tx}{Transmitter}
  \acro{Rx}{Receiver}
  \acro{AI}{Artificial Intelligence}
  \acro{DM}{Diffusion Models}
  \acro{KL}{Kullback-Leibler}
  \acro{BCE}{Binary Cross-Entropy}
  \acro{MIMO}{Multiple-Input Multiple-Output}
  \acro{CSI}{Channel State Information}
  \acro{GMM}{Gaussian Mixture Model}
  \acro{BIC}{Bayesian Information Criterion}
  \acro{SNR}{Signal-to-Noise Ratio}
  \acro{GPU}{Graphics Processing Unit}
  \acro{MEVT}{Multivariate Extreme Value Theory}
  \acro{UEVT}{Univariate Extreme Value Theory}
  \acro{MEL}{Maximum Empirical Likelihood}
  \acro{UAV}{Unmanned Aerial Vehicle}
  \acro{FADE}{\textbf{F}low-based
\textbf{A}ngular \textbf{D}ensity \textbf{E}stimation}
\end{acronym}

\begin{abstract}
Ultra-Reliable Low-Latency Communication (URLLC) is vulnerable to rare low-received-power events, and the spatial-diversity benefits of multiple receivers can be significantly reduced when they experience simultaneous deep fading. Multivariate Extreme Value Theory (MEVT) provides a principled framework for characterizing such joint tail behavior. However, conventional MEVT methods often rely on restrictive parametric assumptions or discrete representations of the dependence structure, which may not fully capture complex multivariate tail behavior. This paper proposes Flow-based Angular Density Estimation (FADE), a flow-based approach for characterizing the dependence between extreme events across receivers within the MEVT framework. FADE transforms multivariate extremes into a standard extreme-value representation and uses a flow-based density estimator to learn the dependence structure from the transformed data without imposing a predefined parametric form. The proposed framework is evaluated using experimentally collected data from a UAV communication system, in which the UAV serves as the transmitter and communicates with two ground receivers. FADE provides closer agreement with the empirical joint CDF, achieving reductions of up to 22.4\% in MAE compared with the state-of-the-art Logistic and Maximum Empirical Likelihood (MEL) methods, while also improving distributional agreement and reducing absolute bias.

\end{abstract}


\section{Introduction}
\label{sec:introduction}

\ac{URLLC} is intended for wireless applications where communication must remain highly reliable, including autonomous systems, industrial automation and remote control \cite{urllc_01}. \ac{URLLC} operates at packet error rates between $10^{-9}$ and $10^{-5}$, with communication delays constrained to only a few milliseconds or less \cite{bennis_tail}. Utilizing multiple antennas can improve reliability through spatial diversity. Therefore, characterizing the joint occurrence of extreme channel observations across receivers is important for ultra-reliable communication.

\ac{EVT} provides tools for studying observations in the tails of a distribution, making it suitable for characterizing rare channel outages \cite{scoles}. Its application to wireless channels has enabled tail-aware channel modeling for \ac{URLLC} \cite{evt_01} and has subsequently been used for EVT-based channel prediction in wireless systems \cite{mehrnia2025evt}. \ac{MEVT} additionally provides tools for modeling the extremes of multiple receivers and describing the dependence between extreme observations across receivers \cite{evt_03}. In this framework, the dependence between extreme observations is characterized by the angular, or spectral, distribution, which describes the relative contribution of each variable to the joint extremes \cite{scoles}. Several approaches have been proposed to model multivariate extreme dependence. The logistic model provides a compact parametric description of the dependence structure but imposes a specific functional form on the dependence between extreme observations \cite{scoles,evt_03}. \ac{MEL} avoids this parametric restriction by estimating the spectral measure directly from extreme observations \cite{mel_mevt}. However, its discrete representation of the spectral measure can limit its scalability and ability to represent complex dependence structures. These limitations motivate a data-driven approach for modeling the dependence structure while retaining the probabilistic formulation required by \ac{MEVT}.

Generative \ac{AI} has recently been investigated for wireless channel modeling and tail estimation \cite{flow_channelest_01,flow_channelest_02,genai_saad,genai_evt_01,genai_evt_02}. Recently, novel \ac{GAN}-based methods have been proposed for wireless channel modeling and tail estimation, including the estimation of \ac{GPD} model parameters for tail modeling from limited measurements and tail-aware channel estimation \cite{genai_evt_01,genai_evt_02}. Normalizing flows provide a generative approach that can learn complex probability distributions while providing an explicit continuous density through invertible transformations \cite{flow_01,flow_02}. Normalizing Flows provide a generative framework for learning complex probability distributions through invertible transformations while providing an explicit continuous density \cite{flow_01,flow_02}. Normalizing Flows have also recently been explored for wireless channel estimation, including MIMO channel recovery using flow matching and recursive flow models \cite{flow_channelest_01,flow_channelest_02}.
 However, these methods do not focus on modeling the joint tail behavior of multiple wireless channels in MIMO systems.

This paper proposes \ac{FADE}, a normalizing-flow framework for multichannel tail estimation within the \ac{MEVT} framework. The main contributions of this paper are summarized as follows. First, a normalizing-flow-based method is developed for modeling the dependence between extreme observations across multiple receivers. Second, the dependence structure is learned from multivariate extreme observations using a continuous density model and incorporated into the \ac{MEVT} formulation to obtain the joint distribution. Finally, the proposed framework is evaluated using experimentally collected \ac{UAV} communication data from a moving UAV transmitter and two ground-based receivers, and its performance is compared with conventional methods for multivariate extreme dependence modeling.

The rest of the paper is organized as follows: Section II presents the system model and assumptions throughout the paper. Section III describes the proposed FADE framework. Section IV presents the experimental setup and numerical results. Finally, Section V concludes the paper and discusses future work.

\section{System Model}
\label{sec:system_model}

We consider a wireless communication system consisting of a single transmitter and $M$ spatially separated receivers that simultaneously observe the same transmitted signal. Let
\[
X_i = \{x_{i,1},x_{i,2},\ldots,x_{i,N}\}
\]
denote the received power observations at the $i$th receiver, where $x_{i,n}$ denotes the received power at the $i$th receiver for the $n$th observation, $i=1,\ldots,M$, and $N$ is the number of observations collected at each receiver.

Low received power levels may lead to outage, defined at the $i$th receiver as
\begin{equation}
\mathcal{O}_i = \{x_{i,n} \in X_i : x_{i,n} \leq u_i\}.
\end{equation}
where $u_{i}$ is the received power outage threshold for receiver $i$. Therefore, the lower tail of the received power distribution at each receiver is important for characterizing individual outage events. Since the receivers observe the same transmitted signal and may be affected by common propagation conditions, extreme low received power events can occur dependently across receivers. This requires the derivation of the joint tail distribution across receivers.

\subsection{Univariate Extreme Value Theory (UEVT)}
\label{subsec:univariate_evt}

For each receiver, we focus on the lower tail of the received power distribution, which corresponds to rare low received power values. Using the \ac{POT} approach, a low threshold $u_i$, corresponding to approximately the 0.1\% quantile of the received power distribution, is selected for the $i$th receiver. Let $Y_i$ denote a generic random variable with the same distribution as the received power observations in $X_i$ at the $i$th receiver. The observations below this threshold are represented by their deficits,

\begin{equation}
Z_i = u_i - Y_i,
\qquad Y_i < u_i,
\end{equation}

where $Z_i \geq 0$ measures how far an observation falls below the threshold. According to the Pickands--Balkema--de Haan theorem \cite{scoles}, the conditional distribution of these deficits can be approximated by a \ac{GPD},

\begin{equation}
G_i(z)
=
1-
\left(
1+\frac{\xi_i z}{\beta_i}
\right)^{-1/\xi_i},
\end{equation}

where $\xi_i$ and $\beta_i>0$ are the shape and scale parameters of \ac{GPD}, respectively, and the support satisfies
$1+\xi_i z/\beta_i>0$.

The fitted \ac{GPD} therefore provides a model for the lower tail behavior of the received power at each receiver. 

\subsection{Multivariate Extreme Value Theory (MEVT)}
\label{subsec:multivariate_evt}

\ac{MEVT} extends the univariate analysis to jointly characterize extreme observations across multiple receivers. To separate the marginal tail behavior from the dependence structure, observations below the selected thresholds are first identified and represented by their deficits. The marginal deficits are then modeled using the \ac{GPD} and transformed to a common standard Fr\'echet scale. The dependence between the transformed deficits is subsequently characterized to obtain the joint CDF of the extreme observations across multiple receivers \cite{scoles,dutfoy2014multivariate,evt_03}.

Let $G_i(\cdot)$ denote the fitted marginal \ac{GPD} for the deficits at the $i$th receiver. The corresponding standard Fr\'echet transformation is given by

\begin{equation}
\label{frechet}
\tilde{Z}_i
=
-\frac{1}{\ln\!\left(G_i(Z_i)\right)},
\qquad
i=1,\ldots,M,
\end{equation}

where $Z_i$ denotes the lower tail exceedances at the $i$th receiver. Each
transformed variable $\tilde{Z}_i$ follows a standard Fr\'echet
distribution with cumulative distribution function

\begin{equation}
P(\tilde{Z}_i\leq\tilde{z}_i)
=
\exp\!\left(-\frac{1}{\tilde{z}_i}\right),
\qquad
\tilde{z}_i>0.
\end{equation}
The transformed observations are collected into the vector $\tilde{\mathbf{Z}}=(\tilde{Z}_1,\tilde{Z}_2,\ldots,\tilde{Z}_M)$. Under the framework of \ac{MEVT}, the joint distribution of $\tilde{\mathbf{Z}}$ is represented by:

\begin{equation}
\begin{aligned}
\label{joint_cdf}
G(\tilde{\mathbf{z}})
&=
P\left(
\tilde{Z}_1\leq\tilde{z}_1,\ldots,
\tilde{Z}_M\leq\tilde{z}_M
\right) \\
&=
P\left(
\tilde{\mathbf{Z}}\leq\tilde{\mathbf{z}}
\right)
=
\exp\!\left[-V(\tilde{\mathbf{z}})\right],
\end{aligned}
\end{equation}
where $V(\tilde{\mathbf{z}})$ is the exponent function \cite{dutfoy2014multivariate}.

To describe the magnitude and relative contribution of each receiver during an extreme event, a pseudo-polar transformation is applied to $\tilde{\mathbf{Z}}$. The transformation is defined as

\begin{equation}
T_p:\mathbb{R}_+^M
\rightarrow
(0,\infty)\times S_M,
\end{equation}

such that $(R,\mathbf{W})=T_p(\tilde{\mathbf{Z}})$, where the radial component $R=\sum_{i=1}^{M}\tilde{Z}_i$ represents the overall magnitude of the extreme observation, and the angular components are given by $W_i=\tilde{Z}_i/R$, $i=1,\ldots,M$, representing the relative contribution of each receiver. By construction, $\sum_{i=1}^{M}W_i=1$, so the angular component $\mathbf{W}$ lies on the $(M-1)$-dimensional unit simplex, denoted by $S_M$. The simplex is defined as

\begin{equation}
S_M
=
\left\{
\mathbf{w}\in[0,1]^M:
\sum_{i=1}^{M}w_i=1
\right\}.
\end{equation}
The inverse transformation is given by $T_p^{-1}(R,\mathbf{W})=R\mathbf{W}$. Let $H$ denote the spectral measure defined on $S_M$. The exponent function in \eqref{joint_cdf} is then given by \cite{ksample_mevt}

\begin{equation}
V(\mathbf{z})
=
M
\int_{S_M}
\max_{1\leq i\leq M}
\left(
\frac{w_i}{z_i}
\right)
H(d\mathbf{w}).
\end{equation}
The spectral measure should satisfy the mean constraint

\begin{equation}
\label{mean_constraint}
\int_{S_M}
w_i\,H(d\mathbf{w})
=
\frac{1}{M},
\qquad
i=1,\ldots,M,
\end{equation}
which ensures that each marginal distribution remains standard Fr\'echet \cite{dutfoy2014multivariate}.

If $H$ is absolutely continuous on $S_M$, it can be represented by an angular density $h(\mathbf{w})$ such that $H(d\mathbf{w})=h(\mathbf{w})\,d\mathbf{w}$. Estimating the angular density $h(\mathbf{w})$ becomes increasingly challenging as the number of receivers $M$ increases. Since $\mathbf{w}$ lies on the $(M-1)$-dimensional simplex, the dimensionality of the density increases with $M$, leading to the curse of dimensionality and computational requirements. Therefore, flexible and scalable density estimation methods are needed for higher dimensional multivariate extreme value models.

\subsection{Normalizing Flows for Density Estimation}
\label{subsec:normalizing_flows}

Normalizing flows are likelihood-based generative models that can learn complex probability distributions through a sequence of invertible and differentiable transformations \cite{flow_01,flow_02}. This provides an explicit formulation of the probability density, making normalizing flows suitable for density estimation. Let $\mathbf{u}\in\mathbb{R}^{D}$ denote a random vector with an unknown density, and let $\mathbf{v}\in\mathbb{R}^{D}$ be a latent vector with a simple base distribution. The two vectors are related through an invertible transformation $\mathbf{u}=f_{\boldsymbol{\theta}}(\mathbf{v})$, where $f_{\boldsymbol{\theta}}$ is a bijective mapping.\footnote{A bijective mapping is a one to one and onto mapping, such that each input has a unique output and each output corresponds to a unique input.} The density of $\mathbf{u}$ can then be obtained using \cite{flow_01,flow_02},

\begin{equation}
\label{change_of_variable}
p_{\mathbf{U}}(\mathbf{u})
=
p_{\mathbf{V}}
\left(
f_{\boldsymbol{\theta}}^{-1}(\mathbf{u})
\right)
\left|
\det
\frac{\partial
f_{\boldsymbol{\theta}}^{-1}(\mathbf{u})}
{\partial\mathbf{u}}
\right|.
\end{equation}

In practice, $f$ is constructed as a composition of multiple invertible transformations to provide sufficient flexibility for modeling complex distributions. The parameters of these transformations are learned by maximizing the likelihood of the observed data. This explicit likelihood formulation makes normalizing flows suitable for complex probability density estimation.

\section{Proposed Flow-based Framework for Multichannel Estimation}
\label{sec:method}

This section presents the proposed \ac{FADE} framework for estimating the angular density
in multivariate extreme value analysis. \ac{FADE} learns a continuous density from the observed angular samples. The framework maps the angular observations from the simplex to an unconstrained space, models their density using an invertible generative model, and transforms the resulting density back to the simplex. The model is optimized using a composite objective that combines likelihood-based density estimation with the mean constraint required by the multivariate extreme value formulation in \eqref{mean_constraint}. The following subsections describe the \ac{FADE}
framework, its network architecture, and the objective
function used for learning.

\subsection{FADE: Flow-based Angular Density Estimation}

Let the observed angular samples be $\{\mathbf{w}_k\}_{k=1}^{K}$, where $\mathbf{w}_k\in S_M$ and $K$ denotes the number of observed mutual extreme events, i.e., events for which $X_i<u_i$ holds simultaneously for all $i=1,\ldots,M$.
The objective is to estimate the angular density $h(\mathbf{w})$. Since the angular samples lie on the constrained simplex $S_M$,
they are first mapped to an unconstrained Euclidean space
through a bijective transformation
\begin{equation}
\label{eq:t}
T:S_M\rightarrow\mathbb{R}^{M-1},
\end{equation}
yielding $\mathbf{u}=T(\mathbf{w})$. A normalizing flow is then used to model the density of
$\mathbf{u}$. Let $\mathbf{v}$ denote a latent variable with
a standard multivariate Gaussian distribution, $\mathbf{v}\sim\mathcal{N}(\mathbf{0},\mathbf{I})$. The flow consists of a sequence of $L$ invertible
transformations. Starting from the latent variable
$\mathbf{v}$, the overall transformation can be represented as

\begin{equation}
\mathbf{q}_0 = \mathbf{v}
\xrightarrow{f_1}
\mathbf{q}_1
\xrightarrow{f_2}
\cdots
\xrightarrow{f_{L-1}}
\mathbf{q}_{L-1}
\xrightarrow{f_L}
\mathbf{q}_L = \mathbf{u}
\xrightarrow{T^{-1}}
\mathbf{w},
\end{equation}
where $\mathbf{q}_l$ denotes the intermediate representation after
the $l$th flow transformation, $l=1,\ldots,L$, and $\mathbf{q}_L$
is the final flow output in the unconstrained Euclidean space. The inverse
transformation $T^{-1}$ then maps the final flow output back to the
simplex. The overall flow transformation is defined as
$f_{\boldsymbol{\theta}} = f_L \circ f_{L-1} \circ \cdots \circ f_1$,
where $\boldsymbol{\theta}$ denotes the flow parameters. Since the density in \eqref{change_of_variable} depends on the
Jacobian determinant of the flow transformation, while the flow is
defined in the forward direction, the two Jacobians are related by

\begin{equation}
\left|
\det
\frac{\partial f_{\boldsymbol{\theta}}^{-1}(\mathbf{u})}
{\partial\mathbf{u}}
\right|
=
\frac{1}{
\left|
\det
\frac{\partial f_{\boldsymbol{\theta}}(\mathbf{v})}
{\partial\mathbf{v}}
\right|
},
\qquad
\mathbf{v}=f_{\boldsymbol{\theta}}^{-1}(\mathbf{u}).
\end{equation}
Therefore, the density of $\mathbf{u}$ is modeled by
\eqref{change_of_variable}, and the log-Jacobian determinant of the
forward flow is computed as

\begin{equation}
\log
\left|
\det
\frac{\partial f_{\boldsymbol{\theta}}}
{\partial\mathbf{v}}
\right|
=
\sum_{l=1}^{L}
\log
\left|
\det
\frac{\partial f_l}
{\partial\mathbf{q}_{l-1}}
\right|,
\end{equation}
where $\mathbf{q}_0=\mathbf{v}$. The angular density is then obtained by transforming the
density back to the simplex,

\begin{equation}
h_{\boldsymbol{\theta}}(\mathbf{w})
=
p_{\mathbf{U}}
\left(
T(\mathbf{w})
\right)
\left|
\det
\frac{\partial T(\mathbf{w})}
{\partial\mathbf{w}}
\right|.
\end{equation}

\subsection{FADE Optimization}
\label{subsec:fade_optimization}

The parameters $\boldsymbol{\theta}$ are learned using a
composite objective consisting of three complementary loss
terms: the negative log-likelihood, the MEVT consistency loss,
and the angular mean constraint loss.
Each term addresses a different requirement of the estimated
angular density.

The negative log-likelihood loss is defined as

\begin{equation}
\mathcal{L}_{\mathrm{NLL}}
=
-\frac{1}{K}
\sum_{k=1}^{K}
\log
h_{\boldsymbol{\theta}}(\mathbf{w}_k),
\end{equation}
where $h_{\boldsymbol{\theta}}(\mathbf{w}_k)$ is the estimated
angular density evaluated at the observed angular sample
$\mathbf{w}_k$. This term encourages the estimated angular density
to assign high probability to the observed mutual extreme samples
and provides the primary likelihood-based learning objective.

The MEVT loss, $\mathcal{L}_{\mathrm{MEVT}}$, is defined to penalize the difference between the
estimated MEVT-based joint distribution and the empirical joint
distribution and is given by

\begin{equation}
\mathcal{L}_{\mathrm{MEVT}}
=
\sqrt{
\frac{1}{K}
\sum_{k=1}^{K}
\left[
G_{\mathrm{emp}}(\mathbf{w}_k)
-
G_{\boldsymbol{\theta}}(\mathbf{w}_k)
\right]^2
}.
\end{equation}
Here, $G_{\mathrm{emp}}$ denotes the empirical joint distribution
and $G_{\boldsymbol{\theta}}$ denotes the joint distribution estimated by MEVT. Thus, $\mathcal{L}_{\mathrm{MEVT}}$ represents the
\ac{RMSE} between the empirical and estimated MEVT-based joint
distributions.

The angular density must also satisfy the mean constraint of
the spectral measure given in \eqref{mean_constraint}. This
requirement is incorporated through the penalty

\begin{equation}
\mathcal{L}_{\mathrm{MC}}
=
\sum_{i=1}^{M}
\left(
\int_{S_M}
w_i h_{\boldsymbol{\theta}}(\mathbf{w})
\,d\mathbf{w}
-
\frac{1}{M}
\right)^2,
\end{equation}
where $w_i$ denotes the $i$th component of the angular vector
$\mathbf{w}$. The total optimization objective is then defined as

\begin{equation}
\mathcal{L}
=
\lambda_{\mathrm{NLL}}\mathcal{L}_{\mathrm{NLL}}
+
\lambda_{\mathrm{MEVT}}\mathcal{L}_{\mathrm{MEVT}}
+
\lambda_{\mathrm{MC}}\mathcal{L}_{\mathrm{MC}},
\label{eq:loss}
\end{equation}
where
$\lambda_{\mathrm{NLL}}$,
$\lambda_{\mathrm{MEVT}}$, and
$\lambda_{\mathrm{MC}}$
are hyperparameters that control the relative contribution of the three loss terms.
\section{Numerical Results}
\label{sec:numerical_results}

This section evaluates the performance of the proposed \ac{FADE} model for joint tail dependence estimation using experimentally collected \ac{UAV} communication data. The measurements consist of a single UAV transmitter and two spatially separated ground receivers. The evaluation first considers the marginal tail behavior of the two receivers using \ac{UEVT} as described in Section~\ref{subsec:univariate_evt}, followed by the transformation of the marginal observations to standard Fr\'echet margins as described in \eqref{frechet}. The joint tail behavior is then modeled using the proposed \ac{FADE} model. For comparison, the same joint tail behavior is also modeled using the state-of-the-art logistic \ac{MEVT} model\cite{evt_03,scoles} and the \ac{MEL} approach\cite{mel_mevt}.

The experimental setup and dataset are first described in Section~\ref{subsec:experimental_setup}, followed by the model architecture and training procedure in Section~\ref{subsec:model_architecture}. The results are then evaluated through the marginal transformation, joint CDF comparison, and quantitative error metrics in Section~\ref{subsec:performance_evaluation}.

\subsection{Experimental Setup and Dataset}
\label{subsec:experimental_setup}

The experimental data were collected using two ground-based receivers
that simultaneously recorded the signal transmitted by a moving
\ac{UAV}. The \ac{UAV} flew at an altitude of approximately 40 m above
ground level and moved along a flight trajectory with speeds of up to
approximately 26 km/h. The transmitted signal was an OFDM waveform
with $N_{\mathrm{FFT}}=160$ subcarriers, of which 64 were active. A
Zadoff--Chu sequence of length 64 was transmitted over the active
subcarriers. The carrier frequency was 3.4 GHz, the RF bandwidth was
2 MHz, the active subcarrier bandwidth was 1 MHz, and the sampling
frequency was 2.5 MHz. The receivers were separated by approximately 18 m, corresponding to approximately 200 wavelengths at 3.4 GHz, providing substantial spatial separation for evaluating the dependence between the received signals. For each received OFDM symbol, the channel frequency response was estimated on the active subcarriers using the known transmitted symbols. The received power was then obtained from the squared magnitude of the estimated channel response and averaged over the active subcarriers, resulting in one received power observation per OFDM symbol.

To reduce the effect of large scale power variations, a moving average
is computed from the received power sequence and subtracted from the
original sequence. The resulting
sequence from each receiver is then divided into non-overlapping blocks of 1000
samples, and the minimum value of each block is retained. Thus,
the resulting samples represent local minimum received power events and
provide an approximately independent sequence, which is important for
the i.i.d. assumption underlying the subsequent extreme value analysis \cite{scoles,evt_01}. For each receiver, the lower-tail threshold below which the received power samples are considered extreme is selected independently at approximately the 0.1\% quantile of the received power distribution. Both receivers resulted in the same selected threshold of $-11$ dBm. The \ac{GPD} is
then fitted to the corresponding threshold exceedances using the
\ac{GAN} model introduced in \cite{genai_evt_01,genai_evt_02}, yielding
the estimated shape and scale parameters. The corresponding estimated
tail parameters for the two receivers are summarized in
Table~\ref{tab:marginal}.

\begin{table}[h]
\centering
\caption{Estimated marginal tail parameters.}
\label{tab:marginal}
\begin{tabular}{c|ccc}
\hline
Receiver & u & $\xi$ & $\beta$ \\
\hline
1 & -11 & -0.1696 & 5.1047\\
2 & -11 & -0.1957 & 5.2135\\
\hline
\end{tabular}
\end{table}

The fitted marginal distributions are used to transform the corresponding
threshold exceedances to standard Fr\'echet margins for subsequent joint
tail dependence analysis. For the joint analysis, only the threshold
exceedances identified as extreme at both receivers, i.e., those for
which both receivers simultaneously fall below their respective
thresholds, are considered. Applying this criterion yields $K=589$ simultaneous extreme events, which are used for the subsequent joint tail dependence analysis.

\subsection{Model Architecture and Training}
\label{subsec:model_architecture}

The joint tail dependence is first modeled using the proposed
\ac{FADE} model. The \ac{FADE} model is implemented using a Masked Piecewise Rational Quadratic Autoregressive Transform (Masked PRQ-AR), following the neural spline flow formulation \cite{flow_04}. The Masked PRQ-AR transformation is configured with $8$ hidden
features, $5$ rational-quadratic bins, linear tails, and a tail
bound of $3$. A standard normal distribution is used as the base
distribution.

For each mutual extreme event, the angular component used for
training is obtained from the standard Fr\'echet observations.
The transformation $T$ in \eqref{eq:t} is selected as the logit
transformation. The flow is trained on these transformed angular samples. For training and evaluation, the $K=589$ simultaneous extreme events are randomly partitioned into training, validation, and test sets containing $509$, $40$, and $40$ observations, respectively. The validation set is used for early stopping, and the test set is reserved for final evaluation. This procedure is repeated over 30 random splits to assess the variability of the estimation errors across data partitions. The weighting coefficients in \eqref{eq:loss} are selected empirically
to balance the contributions of the three loss terms. The selected values are $\lambda_{\mathrm{NLL}}=5$, $\lambda_{\mathrm{MEVT}}=50$, and $\lambda_{\mathrm{MC}}=100$. The flow parameters are optimized using the Adam optimizer with
a learning rate of $10^{-3}$ and a weight decay of $10^{-4}$. The
model is trained for a maximum of $200$ epochs, with early stopping based on the validation set and a patience of 10 epochs.

\subsection{Comparison Baselines}
\label{subsec:comparison_baselines}

\subsubsection{Logistic MEVT}
For the logistic \ac{MEVT} model, the exponent function in \eqref{joint_cdf} is parameterized by $\alpha$ and is given by

\begin{equation}
V_{\mathrm{L}}(\tilde{\mathbf{z}})
=
\left(
\tilde{z}_1^{-1/\alpha} + \tilde{z}_2^{-1/\alpha}
\right)^{\alpha}.
\end{equation}

The parameter $\alpha$ is estimated by maximizing the likelihood
defined by the logistic joint density\cite{scoles,evt_03}. The resulting estimate is
$\alpha=0.6736$, which is then used to obtain
$V_{\mathrm{L}}(\tilde{\mathbf{z}})$ and the corresponding joint
CDF.

\subsubsection{MEL}
For the \ac{MEL} approach, the angular components $\mathbf{w}_k$ are
used to estimate empirical likelihood weights $p_k$. The weights are
obtained by maximizing the empirical likelihood subject to the
normalization constraint and the spectral mean constraint. The
resulting weights should satisfy

\begin{equation}
\sum_{k=1}^{K} p_k = 1,
\qquad
\sum_{k=1}^{K} p_k w_{k}
=
\frac{1}{2}.
\end{equation}
The estimated weights are then used to calculate the exponent
function as

\begin{equation}
V_{\mathrm{MEL}}(\tilde{\mathbf{z}})
=
2\sum_{k=1}^{K}
p_k
\max
\left(
\frac{w_{k}}{\tilde{z}_1},\frac{1-w_{k}}{\tilde{z}_2}
\right),
\end{equation}
which is then used in the general MEVT formulation to obtain the
corresponding joint CDF.

\subsection{Performance Evaluation}
\label{subsec:performance_evaluation}

\begin{figure}[h]
\centering
\includegraphics[width=0.7\linewidth]{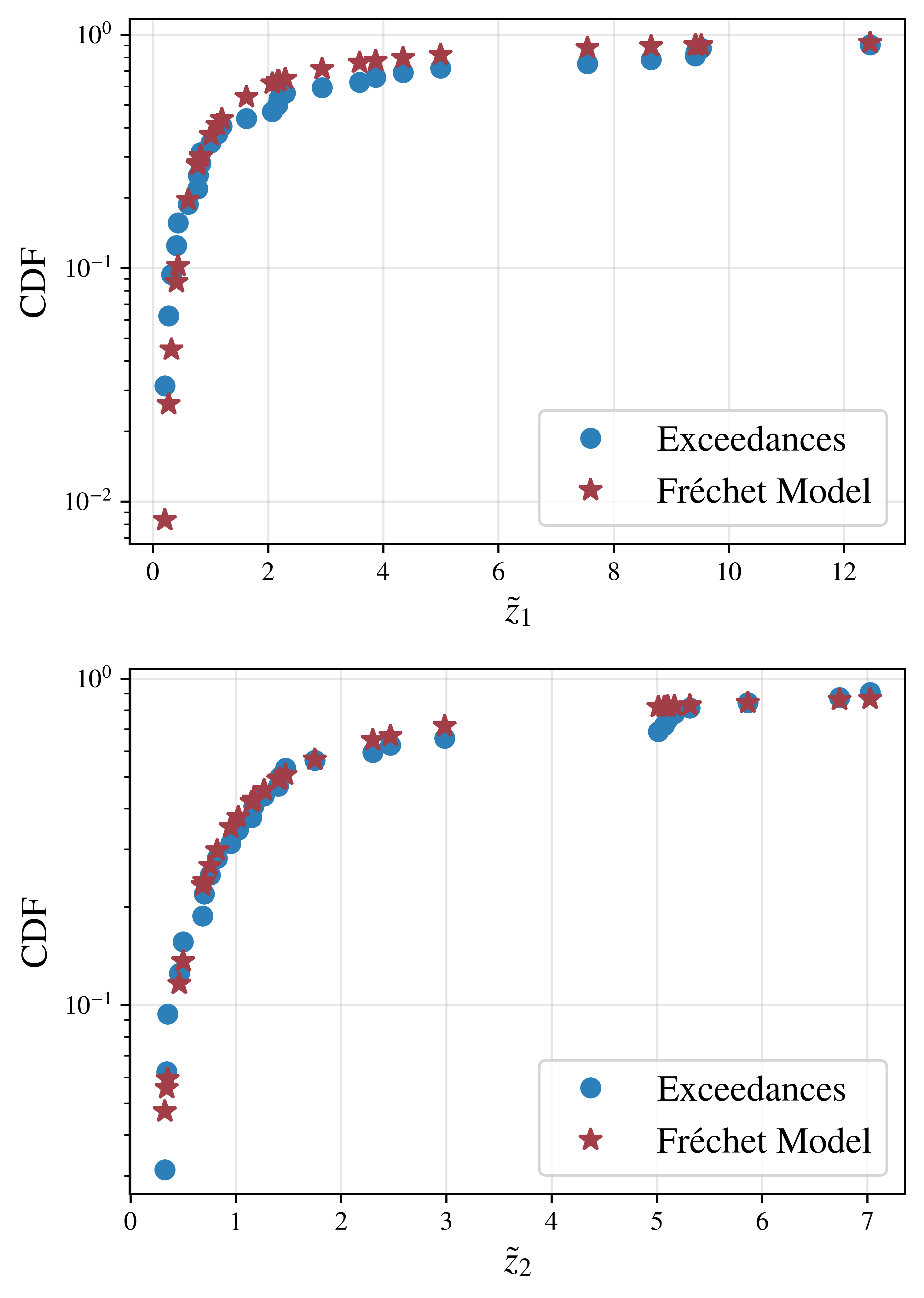}
\caption{Standard Fr\'echet model and transformed exceedances for the two receivers.}
\label{fig:frechet_fit}
\end{figure}

Fig.~\ref{fig:frechet_fit} shows the empirical CDFs of the
transformed tail exceedances $\tilde{Z}_i$ and the theoretical
standard Fr\'echet CDF, $\exp(-1/\tilde{Z}_i)$, for the two receivers.
The close agreement indicates that the fitted marginal \ac{GPD} models
provide a good representation of the tail exceedances and that the
resulting transformation gives appropriate standard Fr\'echet
margins for the subsequent joint tail analysis.

\begin{figure}[h]
\centering
\includegraphics[width=0.85\linewidth]{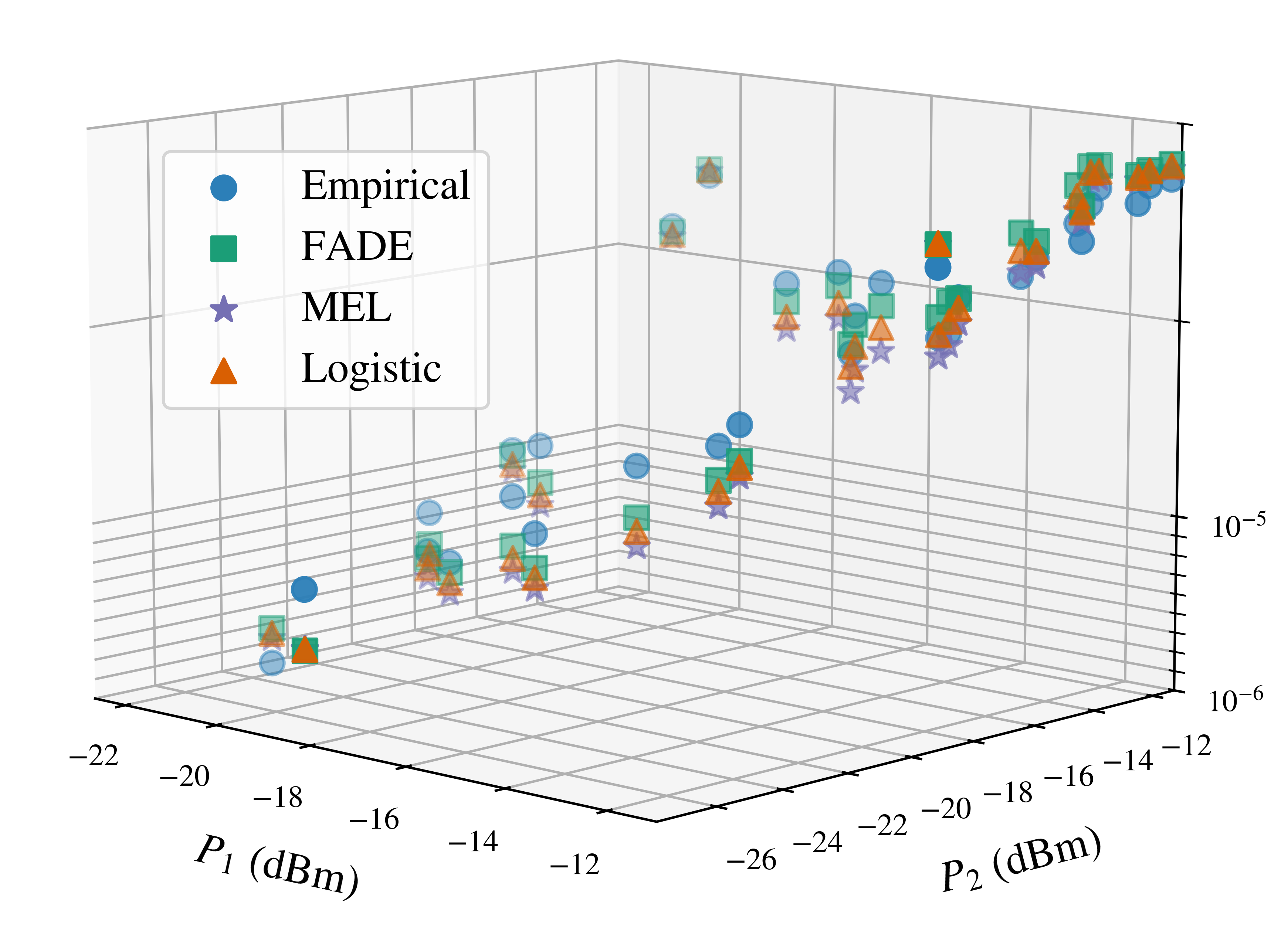}
\caption{Empirical and estimated joint CDFs of the two
receivers for the Logistic, MEL, and FADE models.}
\label{fig:joint_cdf}
\end{figure}

Fig.~\ref{fig:joint_cdf} compares the empirical joint CDF with
the estimates obtained from the Logistic, MEL, and \ac{FADE} models.
FADE shows closer agreement with the empirical CDF, particularly
in the deepest joint tail region, indicating more accurate
estimation of low-probability joint extreme events.

\begin{figure}[h]
\centering
\includegraphics[width=0.85\linewidth]{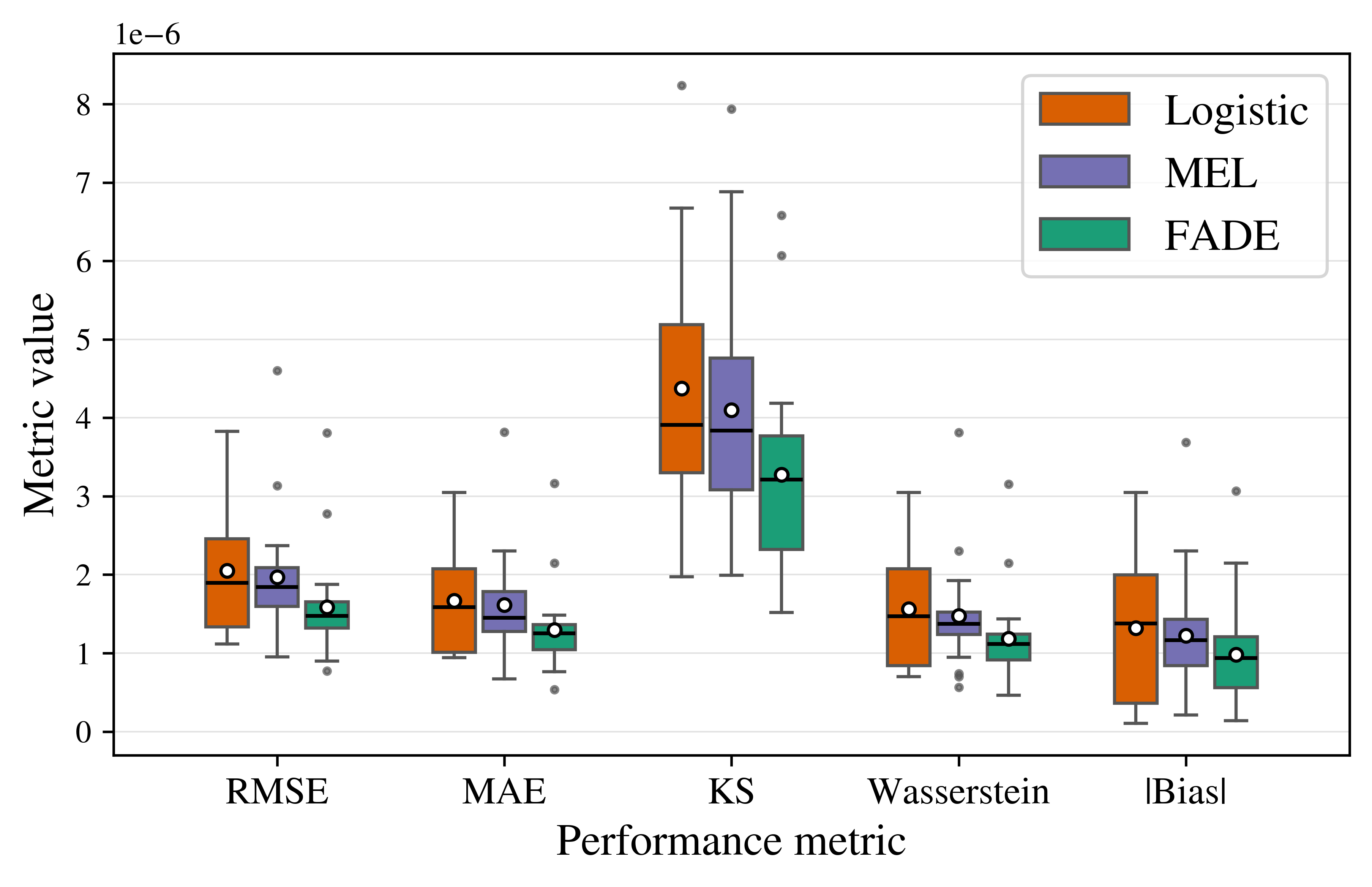}
\caption{Distribution of joint CDF estimation errors across repeated train/test splits for the Logistic, MEL, and FADE models. White markers denote the mean values.}
\label{fig:boxplot}
\end{figure}

Fig.~\ref{fig:boxplot} presents the distributions of the joint CDF estimation errors for the Logistic, \ac{MEL}, and proposed \ac{FADE} models across 30 repeated random data splits, each comprising training, validation, and test sets. The boxplots show the median, interquartile range, and variability of each metric, while the white markers indicate the mean values. Averaged across the train/test splits, \ac{FADE} achieves reductions in \ac{RMSE}, \ac{MAE}, \ac{KS} distance, Wasserstein distance, and absolute bias of 22.69\%, 22.44\%, 25.12\%, 24.37\%, and 25.92\%, respectively, compared with Logistic. The corresponding reductions compared with \ac{MEL} are 19.55\%, 19.76\%, 20.06\%, 19.83\%, and 20.05\%, respectively. The lower \ac{RMSE} and \ac{MAE} indicate improved pointwise agreement with the empirical joint CDF, while the lower \ac{KS} and Wasserstein distances indicate reduced maximum pointwise discrepancies and improved overall distributional agreement, respectively. The lower absolute bias further indicates reduced systematic deviation from the empirical estimates. The consistent improvements across the repeated splits suggest that \ac{FADE} maintains its estimation performance under variations in the training and test data.

\section{Conclusions}
\label{sec:conclusions}

This paper presented a normalizing-flow-based method for modeling the joint tail distribution of received power across multiple receivers. By transforming multivariate extreme observations and learning their dependence structure with a continuous density model, FADE characterizes the joint behavior of extreme received-power events across receivers within the \ac{MEVT} framework. The evaluation on experimentally collected \ac{UAV} data
showed that FADE achieves more accurate joint tail CDF estimation and
lower errors across all evaluated metrics compared with the Logistic
and MEL models. These results demonstrate the effectiveness of FADE
for joint tail modeling of multiple wireless channels. Future work
will extend the evaluation to higher-dimensional multi-receiver
systems and consider different carrier frequencies, UAV trajectories,
and scenarios with both moving transmitters and receivers, to assess the generalizability
of the proposed approach.

\balance 

\ifCLASSOPTIONcaptionsoff
  \newpage
\fi
\bibliographystyle{ieeetr}
\bibliography{GenAI_MEVT_GLOBECOM_Conf}

\end{document}